\documentclass[onecolumn,showpacs]{revtex4}
\usepackage{amssymb}
\usepackage{graphicx}
\usepackage{dcolumn}
\usepackage{bm}

\input epsf
\input{tcilatex}

\begin{document}

\title{{\Large Quasi-Spherical Gravitational Collapse in Any Dimension}}
\author{\textbf{Ujjal Debnath}}
\email{ujjaldebnath@yahoo.com}
\author{\textbf{Subenoy Chakraborty}}
\email{subenoyc@yahoo.co.in}
\affiliation{Department of Mathematics, Jadavpur University, Calcutta-32, India.}
\author{\textbf{John D. Barrow}}
\email{J.D.Barrow@damtp.cam.ac.uk}
\affiliation{DAMTP, Centre for Mathematical Sciences, Cambridge University, Wilberforce
Rd., Cambridge CB3 OWA, UK.}
\date{\today}

\begin{abstract}
We study the occurrence and nature of naked singularities for a dust model
with non-zero cosmological constant in ($n+2$)-dimensional Szekeres
space-times (which possess no Killing vectors) for $n\geq 2$. We find that
central shell-focusing singularities may be locally naked in higher
dimensions but depend sensitively on the choice of initial data. In fact,
the nature of the initial density determines the possibility of naked
singularity in space-times with more than five dimensions. The results are
similar to the collapse in spherically symmetric Tolman-Bondi-Lema\^{\i}tre
space-times.
\end{abstract}

\pacs{0420D, 0420J, 0470B}
\maketitle

\section{{\protect\normalsize \textbf{Introduction }}}

An extensive study [1-6] of gravitational collapse has been carried out of
Tolman-Bondi-Lema\^{\i}tre (TBL) spherically symmetric space-times
containing irrotational dust. Due to simplifications introduced by the
spherical symmetry several generalizations of this model have been
considered. A general conclusion from these studies is that a central
curvature singularity forms but its local or global visibility depends on
the initial data. Also the study of higher-dimensional spherical collapse
reveals the interesting feature that visibility of singularity is impossible
in space-times with more than five dimensions with proper choice of regular
initial data [7-9].\newline

By contrast, there is very little progress in studying non-spherical
collapse. The basic difficulty is the ambiguity of horizon formation in
non-spherical geometries and the influence of gravitational radiation.
Schoen and Yau [10] proposed a sufficient criterion for the formation of
trapped surfaces in an arbitrary space-time but it fails to say anything
about the conditions which lead to the formation of naked singularities.
This problem has been restated by Thorne [11] in the form of a conjecture
known as the \textit{hoop conjecture} which states that \textit{%
\textquotedblleft horizon form when and only when a gravitational mass $M$
gets compacted into a region whose circumference in every direction is $%
C\lesssim 4\pi M$\textquotedblright }. Subsequently, there were attempts
namely numerical simulations of prolate and oblate collapse [12],
gravitational radiation emission in aspherical collapse [13], analytical
studies of prolate collapsing spheroids [14] and others [15,16] to prove or
disprove the conjecture. Interestingly all of them either confirmed or
failed to refute the conjecture.\newline

The quasi-spherical dust collapse models, given by Szekeres metric [17] were
analyzed by Szekeres himself [18], Joshi and Krolak [19], Deshingkar, Joshi
and Jhingan [20] and extensively by Goncalves [21]. The solutions for dust
and a non-zero cosmological constant were found by Barrow and Stein-Schabes
[26]. In this work, we study the gravitational collapse in the recently
generalized ($n+2$)-dimensional Szekeres metric [22]. As in four-dimensional
space-time, this higher dimensional model does not admit any Killing vector
and the description quasi-spherical arises because it has invariant family
of spherical hypersurfaces. The paper is organized as follows: In section II
we derive the basic equations and regularity conditions. In section III we
investigate the formation and the nature of central singularity. We study
the formation of an apparent horizon due to collapse in section IV. The
nature and the strength of the singularity is investigated by an analysis of
the geodesics in sections V and VI respectively. Finally the paper ends with
a short discussion.\newline

\section{{\protect\normalsize \textbf{Basic Equations and regularity
conditions}}}

Recently, dust solutions have been obtained for ($n+2$)-dimensional
Szekeres' space-time metric for which the line element is [22]

\begin{equation}
ds^{2}=dt^{2}-e^{2\alpha }dr^{2}-e^{2\beta }\sum_{i=1}^{n}dx_{i}^{2}
\end{equation}%
where $\alpha $ and $\beta $ are functions of all the ($n+2$) space-time
coordinates. Under the assumption that $\beta ^{\prime }~(=\frac{\partial
\beta }{\partial r})\neq 0$, the explicit form of the metric coefficients
are [22]

\begin{equation}
e^{\beta}=R(t,r)~e^{\nu(r,x_{1},...,x_{n})}
\end{equation}

and

\begin{equation}
e^{\alpha}=\frac{R^{\prime}+R~\nu^{\prime}}{\sqrt{1+f(r)}}
\end{equation}

where

\begin{equation}
e^{-\nu}=A(r)\sum_{i=1}^{n}x_{i}^{2}+\sum_{i=1}^{n}B_{i}(r)x_{i}+C(r)
\end{equation}

and $R$ satisfied the differential equation

\begin{equation}
\dot{R}^{2}=f(r)+\frac{F(r)}{R^{n-1}}+\frac{2\Lambda}{n(n+1)}~R^{2}~.
\end{equation}

Here $\Lambda $ is cosmological constant, $f(r)$ and $F(r)$ are arbitrary
functions of $r$ alone; and the other arbitrary functions, namely $%
A(r),~B_{i}(r)$ and $C(r),$ are restricted by the algebraic relation [22]

\begin{equation}
\sum_{i=1}^{n}B_{i}^{2}-4AC=-1
\end{equation}

The $r$-dependence of these arbitrary functions $A,~B_{i}$ and $C$ plays an
important role in characterizing the geometry of the ($n+1$)-dimensional
space. In particular, if we choose $A(r)=C(r)=\frac{1}{2}$ and $B_{i}(r)=0$ (%
$\forall ~i=1,2,...,n$) then using the transformation [22]
\[
\begin{array}{llll}
x_{1}=Sin\theta _{n}Sin\theta _{n-1}...~~...Sin\theta _{2}Cot\frac{1}{2}%
\theta _{1} &  &  &  \\
&  &  &  \\
x_{2}=Cos\theta _{n}Sin\theta _{n-1}...~~...Sin\theta _{2}Cot\frac{1}{2}%
\theta _{1} &  &  &  \\
&  &  &  \\
x_{3}=Cos\theta _{n-1}Sin\theta _{n-2}...~~...Sin\theta _{2}Cot\frac{1}{2}%
\theta _{1} &  &  &  \\
&  &  &  \\
....~~...~~...~~...~~...~~...~~...~~... &  &  &  \\
&  &  &  \\
x_{n-1}=Cos\theta _{3}Sin\theta _{2}Cot\frac{1}{2}\theta _{1} &  &  &  \\
&  &  &  \\
x_{n}=Cos\theta _{2}Cot\frac{1}{2}\theta _{1} &  &  &
\end{array}%
\]%
\newline
the space-time metric (1) reduces to the spherically symmetric TBL form:

\begin{equation}
ds^{2}=dt^{2}-\frac{R^{\prime }{}^{2}}{1+f(r)}dr^{2}-R^{2}d\Omega _{n}^{2}.
\end{equation}

In the subsequent discussion we shall restrict ourselves to the
quasi-spherical space-time which is characterized by the $r$ dependence of
the function $\nu $ (i.e., $\nu ^{\prime }\neq 0$). From the Einstein field
equations we have an expression for energy density for the dust, as

\begin{equation}
\rho(t,r,x_{1},...,x_{n}) =\frac{n}{2}~\frac{F^{\prime }+(n+1)F\nu ^{\prime }%
}{R^{n}(R^{\prime }+R\nu ^{\prime })}.
\end{equation}

Thus a singularity will occur when either (i) $R=0$ i.e., $\beta =-\infty $
or (ii) $\alpha =-\infty $. Using the standard terminology for spherical
collapse, the first case corresponds to a shell-focusing singularity, while
the second case gives rise to a shell-crossing singularity. As in a TBL
space-time, the shell-crossing singularities are gravitationally weak, and
we shall not consider them any further here. Hence, we shall restrict
ourselves to the situation with $\alpha >-\infty $.\newline

Suppose that $t=t_{i}$ is the initial hypersurface from which the collapse
develops. For the initial data we assume that $R(t_{i},r)$ is a
monotonically increasing function of $r$. So, without any loss of
generality, we can label the dust shells by the choice $R(t_{i},r)=r$. As a
result, the expression for the initial density distribution is given by

\begin{equation}
\rho_{i}(r,x_{1},...,x_{n})=\rho(t_{i},r,x_{1},...,x_{n})=\frac{n}{2}~\frac{%
F^{\prime}+(n+1)F\nu^{\prime}}{r^{n}(1+r\nu^{\prime})}
\end{equation}

If \ we started the collapse from a regular initial hypersurface the
function $\rho _{i}$ must be non-singular (and also positive for a
physically realistic model). Furthermore, in order for the space-time to be
locally flat near $r=0,$ we must have $f(r)\rightarrow 0$ as $r\rightarrow 0$%
. Then, from equation (5), the boundedness of $\dot{R}^{2}$ as $r\rightarrow
0$ demands that $F(r)\sim O(r^{m})$ where $m\geq n-1$. But, for small $r$, $%
\rho _{i}(r)\simeq \frac{nF^{\prime }}{2r^{n}}$ and consequently, for
regular $\rho _{i}(r)$ near $r=0$, we must have $F(r)\sim O(r^{n+1})$. Thus,
starting with a regular initial hypersurface, we can express $F(r)$ and $%
\rho _{i}(r)$ by power series near $r=0$ as

\begin{equation}
F(r)=\sum_{j=0}^{\infty }F_{j}~r^{n+j+1}
\end{equation}%
and

\begin{equation}
\rho _{i}(r)=\sum_{j=0}^{\infty }\rho _{j}~r^{j}.
\end{equation}

As $\nu ^{\prime }$ appears in the expression for the density as well as in
the metric coefficient, we can write

\begin{equation}
\nu ^{\prime }(r)=\sum_{j=-1}^{\infty }\nu _{j}~r^{j}
\end{equation}%
where $\nu _{_{-1}}>-1$.\newline

Now, using these series expansions in equation (9) we have the following
relations between the coefficients,

\begin{eqnarray*}
\rho_{0}=\frac{n(n+1)}{2}F_{0},~~\rho_{1}=\frac{n}{2}\left(n+1+\frac{1}{1+
\nu_{_{-1}}}\right)F_{1},
\end{eqnarray*}
\vspace{-5mm}

\begin{eqnarray*}
\rho_{2}=\frac{n}{2}\left[\left(n+1+\frac{2}{1+ \nu_{_{-1}}}\right)F_{2}-%
\frac{F_{1}\nu_{_{0}}}{(1+\nu_{_{-1}})^{2}}\right],
\end{eqnarray*}
\vspace{-5mm}

\begin{equation}
\rho _{3}=\frac{n}{2}\left[ \left( n+1+\frac{3}{1+\nu _{_{-1}}}\right) F_{3}-%
\frac{2F_{2}\nu _{_{0}}}{(1+\nu _{_{-1}})^{2}}-\frac{(1+\nu _{_{-1}})\nu
_{_{1}}-\nu _{_{0}}^{2}}{(1+\nu _{_{-1}})^{3}}F_{1}\right]
\end{equation}%
and so on.\newline

Finally, if we assume that the density gradient is negative and falls off
rapidly to zero near the centre then we must have $\rho _{1}=0$ and $\rho
_{2}<0$. Consequently, we have the restrictions that $F_{1}=0$ and $F_{2}<0$.

\section{{\protect\normalsize \textbf{Formation of singularity and its nature%
}}}

In order to form a singularity from the gravitational collapse of dust we
first require that all portions of the dust cloud are collapsing i.e., $\dot{%
R}\leq 0$. Let us define $t_{sf}(r)$ and $t_{sc}(r)$ as the time for
shell-focusing and shell-crossing singularities to occur occurring at radial
coordinate $r$. This gives the relations [18]

\begin{equation}
R(t_{sf},r)=0
\end{equation}%
and

\begin{equation}
R^{\prime }(t_{sc},r)+R(t_{sc},r)\nu ^{\prime }(r,x_{1},x_{2},...,x_{n})=0.
\end{equation}%
Note that `$t_{sc}$' may also depend on $x_{1},x_{2},...,x_{n}$.\newline

As mentioned earlier, the shell-crossing singularity is not of much physical
interest so we shall just consider the shell-focusing singularity for the
following two cases:\newline

(i) \textit{Marginally bound case} : $f(r)=0$ with $\Lambda\ne 0$\newline

In this case equation (5) can easily be integrated to give

\begin{equation}
t=t_{i}+\sqrt{\frac{2n}{(n+1)\Lambda}}\left[Sinh^{-1}\left(\sqrt{\frac{%
2\Lambda r^{n+1}}{n(n+1)F(r)}}\right)-Sinh^{-1}\left(\sqrt{\frac{2\Lambda
R^{n+1}} {n(n+1)F(r)}}\right)\right]
\end{equation}%
Also, from equations (14) and (16), we have

\begin{equation}
t_{sf}=t_{i}+\sqrt{\frac{2n}{(n+1)\Lambda}}~Sinh^{-1}\left(\sqrt{\frac{%
2\Lambda r^{n+1}}{n(n+1)F(r)}}\right)
\end{equation}

\bigskip

(ii) \textit{Non-marginally bound case with time symmetry} : $f(r)\neq 0,~%
\dot{R}(t_{i},r)=0, \Lambda=0$\newline

Here, the solution of equation (5) gives $t$ as a function of $r$ :

\begin{equation}
t=t_{i}+\frac{2}{(n+1)\sqrt{F}}\left[\frac{\sqrt{\pi}~\Gamma(b+1)}{%
\Gamma(b+1/2)}~ r^{\frac{n+1}{2}}-R^{\frac{n+1}{2}}~_{2}F_{1}[\frac{1}{2}%
,b,b+1,\left(\frac{R}{r}\right)^{n-1}] \right]
\end{equation}

and so the time of the shell-focusing singularity is given by

\begin{equation}
t_{sf}=t_{i}+\frac{2\sqrt{\pi }}{(n+1)\sqrt{F}}\frac{\Gamma (b+1)}{\Gamma
(b+1/2)}~r^{\frac{n+1}{2}},
\end{equation}%
where $_{2}F_{1}$ is the usual hypergeometric function with $b=\frac{1}{2}+%
\frac{1}{n-1}$~. However, for the five-dimensional space-time ($n=3)$, $R$
has the particularly simple form

\begin{equation}
R^{2}=r^{2}-\frac{F(r)}{r^{2}}(t-t_{i})^{2},
\end{equation}%
and therefore
\begin{equation}
t_{sf}=t_{i}+\frac{r^{2}}{\sqrt{F(r)}}.
\end{equation}

\section{{\protect\normalsize \textbf{Formation of trapped surfaces}}}

The event horizon of observers at infinity plays an important role in the
nature of the singularity. However, due to the complexity of the
calculation, we shall consider a trapped surface which is a compact
space-time 2-surface whose normals on both sides are future-pointing
converging null-geodesic families. In particular, if ($r=$ constant, $t=$
constant) the 2-surface $S_{r,t}$ is a trapped surface then it, and its
entire future development, lie behind the event horizon provided the density
falls off fast enough at infinity.\newline

If $t_{ah}$ is the instant of the formation of apparent horizon then we must
have [7,8,18]

\begin{equation}
2\Lambda R^{n+1}(t_{ah},r)-n(n+1)R^{n-1}(t_{ah},r)+n(n+1)F(r)=0.
\end{equation}%
Thus, for the above two cases, the explicit expressions for $t_{ah}$ are the
following:

\begin{equation}
t=t_{i}+\sqrt{\frac{2n}{(n+1)\Lambda}}\left[Sinh^{-1}\left(\sqrt{\frac{%
2\Lambda r^{n+1}}{n(n+1)F(r)}}\right)-Sinh^{-1}\left(\sqrt{\frac{2\Lambda
R^{n+1}(t_{ah},r)} {n(n+1)F(r)}}\right)\right]
\end{equation}%
for $f(r)=0, \Lambda\ne 0$ and

\begin{equation}
t_{ah}=t_{i}+\frac{2}{(n+1)\sqrt{F}}\left[ \frac{\sqrt{\pi }~\Gamma (b+1)}{%
\Gamma (b+1/2)}~r^{\frac{n+1}{2}}-F^{b}~~_{2}F_{1}[\frac{1}{2},b,b+1,\frac{F%
}{r^{n-1}}]\right]
\end{equation}%
for $f(r)\neq 0,~\dot{R}(t_{i},r)=0, \Lambda=0$ and for $n=3$ we get

\begin{equation}
t_{ah}=t_{i}+r\sqrt{\frac{r^{2}}{F}-1}.
\end{equation}

From the expressions for $t_{sf}$ and $t_{ah},$ we note that the
shell-focusing singularity that appears at $r>0$ is in the future of the
apparent horizon in both cases. But since we are interested in the central
shell-focusing singularity (at $r=0$) we require the time of occurrence of
central shell-focusing singularity $t_{0}(=t_{sf}(0))$. From equations (16),
(19) and (22), taking the limit as $r\rightarrow 0$, we have that
\begin{eqnarray*}
t_{ah}(r)-t_{0}=-\frac{1}{(n+1)F_{0}^{3/2}\sqrt{1+\frac{2\Lambda}{n(n+1)F_{0}%
}}}\left[F_{1}r+ \left(F_{2}-\frac{(4\Lambda+3n(n+1)F_{0})F_{1}^{2}}{%
4F_{0}(2\Lambda+n(n+1)F_{0})}\right)r^{2} +\right.
\end{eqnarray*}
\vspace{-6mm}

\begin{eqnarray*}
\left. \left(F_{3}-\frac{(4\Lambda+3n(n+1)F_{0})F_{1}F_{2}}{%
4F_{0}(2\Lambda+n(n+1)F_{0})}+ \frac{(32\Lambda^{2}+40n(n+1)\Lambda
F_{0}+15n^{2}(n+1)^{2}+F_{0}^{2})F_{1}^{3}}{24F_{0}^{2}(2%
\Lambda+n(n+1)F_{0})^{2}} \right)r^{3}\right.
\end{eqnarray*}
\vspace{-6mm}

\begin{equation}
\left.+......\frac{}{}\right]-\frac{2}{(n+1)F_{0}^{\frac{n-2}{n-1}}}\left[%
F_{0}r^{\frac{n+1}{n-1}}+\frac{1}{n-1}F_{1} r^{\frac{2n}{n-1}}+\frac{1}{n-1}%
\left(F_{2}-\frac{n-2}{2(n-1)}\frac{F_{1}}{F_{0}} \right)r^{\frac{3n-1}{n-1}%
}+...\right]
\end{equation}
for $f(r)=0, \Lambda\ne 0$ and
\begin{eqnarray*}
t_{ah}(r)-t_{0}=-\frac{\sqrt{\pi}}{(n+1)F_{0}^{3/2}}\frac{\Gamma(b+1)}{%
\Gamma(b+1/2)} \left[F_{1}r+\left(F_{2}-\frac{3F_{1}^{2}}{4F_{0}}%
\right)r^{2}+\left(F_{3}- \frac{9F_{1}F_{2}}{4F_{0}}\right.\right.
\end{eqnarray*}
\vspace{-5mm}

\begin{equation}
\left. \left. +\frac{5F_{1}^{3}}{8F_{0}^{2}}\right) r^{3}+...\right] -\frac{2%
}{(n+1)F_{0}^{\frac{n-2}{n-1}}}\left[ F_{0}r^{\frac{n+1}{n-1}}+\frac{1}{n-1}%
F_{1}r^{\frac{2n}{n-1}}+...\right]
\end{equation}

for $f(r)\ne 0,~ \dot{R}(t_{i},r)=0, \Lambda=0$ and \newline

for $n=3$ (with $f(r)\ne 0,~ \dot{R}(t_{i},r)=0, \Lambda=0$)
\begin{equation}
t_{ah}(r)-t_{0}=-\frac{1}{2F_{0}^{3/2}}\left[F_{1}~r+\left(F_{2}+F_{0}^{2}-
\frac{3F_{1}^{2}}{4F_{0}}\right)r^{2}+......\right]
\end{equation}

Here, $t_{0}$ is the time at which the singularity is formed at $r=0,$ and $%
t_{ah}(r)$ is the instant at which a trapped surface is formed at a distance
$r.$ Therefore, if $t_{ah}(r)>t_{0},$ a trapped surface will form later than
the epoch at which any light signal from the singularity can reach an
observer. Hence the necessary condition for a naked singularity to form is $%
t_{ah}(r)>t_{0};$ while that for a black hole to form is
$t_{ah}(r)\geq t_{0} $. It is to be noted that this criteria for
naked singularity is purely local. Hence, in the present problem
it possible to have local naked singularity or a black hole form
under the conditions shown in the Table I.\\\\

\[
\text {TABLE-I}
\]
\[
\begin{tabular}{|l|l|r|r|r|}
\hline\hline
& \multicolumn{1}{c|}{\emph{Naked Singularity}} & \multicolumn{1}{c|}{\emph{%
Black Hole}} &  &  \\ \hline
&  &  &  &  \\
Marginally bound case: & (i) $\rho_{1}<0,~~ \forall n$ & (i) $\rho_{1}=0,
\rho_{2}<0,n=3,$~~~~~~~~~~~~~~~~ &  &  \\
$f(r)=0, \Lambda\ne 0$ & (ii) $\rho_{1}=0, \rho_{2}<0, n=2$ & $%
F_{2}\ge-2F_{0}^{2}\sqrt{1+\frac{\Lambda}{6F_{0}}}$~~~~~~~~~~~~~~~~ &  &  \\
& (iii) $\rho_{1}=0, \rho_{2}<0,n=3,$ & (ii) $\rho_{1}=0, \rho_{2}<0, n\ge4$
~~~~~~~~~~~~~~~ &  &  \\
& ~~~~~$F_{2}<-2F_{0}^{2}\sqrt{1+\frac{\Lambda}{6F_{0}}}$ & (iii) $%
\rho_{1}=0, \rho_{2}=0,\rho_{3}<0,n=2,$~~~~~ &  &  \\
& (iv) $\rho_{1}=0, \rho_{2}=0,\rho_{3}<0,$ & $F_{3}\ge-2F_{0}^{5/2}\sqrt{1+%
\frac{\Lambda}{3F_{0}}}$~~~~~~~~~~~~~ &  &  \\
& ~~~~~$n=2,$ & (iv) $\rho_{1}=0, \rho_{2}=0,\rho_{3}<0, n\ge 3$ ~~~~~ &  &
\\
& ~~~$F_{3}<-2F_{0}^{5/2}\sqrt{1+\frac{\Lambda}{3F_{0}}}$ & (v) $\rho_{1}=0,
\rho_{2}=0,...,\rho_{j}=0,$~~~~~~~~~~~ &  &  \\
&  & $\rho_{j+1}<0, j\ge 3, \forall n$~~~~~~~~~~~~~~~~~~~ &  &  \\ \hline
&  &  &  &  \\
Non-marginally bound & (i) $\rho_{1}<0, \forall n $ & (i) $\rho_{1}=0,
\rho_{2}<0, n=3,F_{2}\ge-F_{0}^{2}$~ &  &  \\
case with time symmetry: & (ii) $\rho_{1}=0, \rho_{2}<0, n=2$ & (ii) $%
\rho_{1}=0, \rho_{2}<0, n\ge 4$~~~~~~~~~~~~~~~ &  &  \\
$f(r)\ne 0,~R(t_{i},r)=0,$ & (iii) $\rho_{1}=0, \rho_{2}<0, n=3,$ & (iii) $%
\rho_{1}=0, \rho_{2}=0,\rho_{3}<0, n=2,$~~~~ &  &  \\
$\Lambda=0$ & ~~~~~$F_{2}<-F_{0}^{2}$ & $F_{3}\ge -\frac{8}{3\pi}F_{0}^{5/2}$%
~~~~~~~~~~~~~~~~~~~~~~~~ &  &  \\
& (iv) $\rho_{1}=0, \rho_{2}=0,\rho_{3}<0,$ & (iv) $\rho_{1}=0,
\rho_{2}=0,\rho_{3}<0, n\ge 3$~~~~~~ &  &  \\
& ~~~~~$n=2, F_{3}<-\frac{8}{3\pi}F_{0}^{5/2}$ & (v) $\rho_{1}=0,
\rho_{2}=0,...,\rho_{j}=0,$ ~~~~~~~~~~ &  &  \\
&  & $\rho_{j+1}<0, j\ge 3, \forall n$~~~~~~~~~~~~~~~~~~~ &  &  \\
\hline\hline
\end{tabular}%
\]%
\newline

However, if we relax the condition on the density gradient (namely the
condition that $\rho _{1}=0,~\rho _{2}<0$ i.e., $F_{1}=0,~F_{2}<0$) then it
is possible to have a naked singularity in all dimensions. These results are
identical to those obtained in the presence of spherical symmetry. Hence the
local nature of singularity is not affected by the geometry of the
space-time with respect to whether it is spherical or quasi-spherical.
However, we note that the Szekeres space-times retain certain special
geometrical properties shared by spherical space-times in all dimensions, in
particular they do not allow gravitational radiation to be present in the
space-time.

\section{{\protect\normalsize \textbf{Geodesics and the nature of singularity%
}}}

Here, for simplicity, we shall consider only the marginally bound case
(i.e., $f(r)=0$) with $\Lambda=0$. Then $R(t,r)$ has the simple explicit
solution (choosing the initial time $t_{i}=0$):

\begin{equation}
R=\left[r^{\frac{n+1}{2}}-\frac{n+1}{2}\sqrt{F(r)}~t\right]^{\frac{2}{n+1}}
\end{equation}

We now follow the geodesic analysis of Joshi and Dwivedi [23] for
TBL model. For simplicity of calculation we introduce the
following functions:
\begin{equation}
\left.
\begin{array}{llll}
X=\frac{R}{r^{a}} &  &  &  \\
&  &  &  \\
\xi =\frac{rF^{\prime }}{F} &  &  &  \\
&  &  &  \\
\eta =\frac{rQ^{\prime }}{Q} &  &  &  \\
&  &  &  \\
\zeta =\frac{F}{r^{a(n-1)}} &  &  &  \\
&  &  &  \\
Q=e^{-\nu } &  &  &  \\
&  &  &  \\
\Theta =\frac{1-\frac{\xi }{n+1}}{r^{\frac{(a-1)(n+1)}{2}}} &  &  &  \\
&  &  &  \\
&  &  &
\end{array}%
\right\}
\end{equation}%
where the constant $a$ is restricted by $a\geq 1$. In order to determine the
nature of the singularity we examine whether it is possible to have outgoing
null geodesics which are terminated in the past at the central singularity $%
r=0$. Suppose this occurs at singularity time $t=t_{0}$ at which $%
R(t_{0},0)=0$. In order to decide this we start with radial null geodesic,
given by

\begin{equation}
\frac{dt}{dr}=e^{\alpha }=\frac{R^{\prime }Q-RQ^{\prime }}{Q}.
\end{equation}%
Next, we introduce the notation $u=r^{a}$, so we write

\begin{equation}
\frac{dR}{du}=U(X,u),
\end{equation}%
where ~~~~$U(X,u)=\left( 1-\sqrt{\frac{\zeta }{X^{n-1}}}\right) \frac{H}{a}+%
\frac{\eta }{a}\sqrt{\frac{\zeta }{X^{n-3}}}$ ~~~~ with~~~~ $H=\frac{\xi }{%
n+1}~X+\frac{\Theta }{X^{\frac{n-3}{2}}}$ .\newline

We now study the limiting behaviour of the function $X$ as we approach the
singularity at $R=0,u=0$ along the radial null geodesic identified above. If
we denote the limiting value by $X_{0}$ then

\begin{eqnarray}
\begin{array}{c}
X_{0}~= \\
{}%
\end{array}
\begin{array}{c}
lim~~~~~~~~~ \frac{R}{u} \\
R\rightarrow 0~ u\rightarrow 0%
\end{array}
\begin{array}{c}
=~lim~~~~~~~~~ \frac{dR}{du} \\
~~~R\rightarrow 0~ u\rightarrow 0%
\end{array}
\begin{array}{c}
=~~lim~~~~~~~~ U(X,u) \\
R\rightarrow 0~ u\rightarrow 0%
\end{array}
\begin{array}{c}
=U(X_{0},0) \\
{}%
\end{array}%
\end{eqnarray}

Thus if this polynomial equation in $X_{0}$ has at least one positive real
root then it is possible to have a radial null geodesic outgoing from the
central singularity. More explicitly, $X_{0}$ is the root of the following
equation in $X$:

\begin{equation}
\left( a-\frac{\xi _{0}}{n+1}\right) X^{n}+\left( \frac{\xi _{0}}{n+1}-\eta
_{0}\right) \sqrt{\zeta _{0}}~X^{\frac{n+1}{2}}-\Theta _{0}X^{\frac{n-1}{2}}+%
\sqrt{\zeta _{0}}~\Theta _{0}=0
\end{equation}%
where suffix `$o$' for the variables $\xi ,\eta ,\zeta $ and $\Theta $
stands for the values of these quantities at $r=0$. As exact analytic
solution is not possible for $X$ so we study the roots by numerical methods.
Table II shows the dependence of the nature of the roots on the variation of
the parameters involved.\newline
\newline
Table II: Positive roots ($X_{0}$) of the eqn. (34) for different
values of the parameters namely $a,\eta _{0},\xi _{0},\zeta
_{0},\Theta _{0}$ and in different dimensions, $n$.

\begin{center}
\begin{tabular}{|l|}
\hline\hline
~$a$~~~~~$\eta_{0}$~~~~~$\xi_{0}$~~~~$\zeta_{0}$~~~~$\Theta_{0}$%
~~~~~~~~~~~~~~~~~~~~~~~~~~~Positive roots ($X_{0}$) \\ \hline
\\
~~~~~~~~~~~~~~~~~~~~~~~~~~~~~~~~~~~~~~4D~~~~~~~~5D~~~~~~~~~6D~~~~~~~~~7D~~~~~~~~8D~~~~~~~~10D~~~~~~14D
\\ \hline\hline
\\
~1~~~~-6~~~~~.05~~~.01~~~~-5~~~~.00997~~~~.09864~~~~.21131~~~~.30963~~~~.38979~~~~.50797~~~.64774
\\
\\
~2~~~~~2~~~~~~~5~~~~~3~~~~~2~~~~3.3019,~~~~1.732,~~~~1.4422,~~~1.1968,~~~~1.2457,~~~1.1699,~~~1.105,
\\
~~~~~~~~~~~~~~~~~~~~~~~~~~~~~~~~~~~3~~~~~~~~~~~1.633~~~~~1.3195~~~~1.3161~~~~~1.1345~~~~1.075~~~~~1.0334
\\
\\
~3~~~~~2~~~~~~~5~~~~~3~~~~~2~~~~~~~$-$~~~~~~~~~~$-$~~~~~~~~~~$-$~~~~~~~~~~$-$%
~~~~~~~~~~~$-$~~~~~~~~~$-$~~~~~~~~~$-$ \\
\\
~5~~~~~4~~~~~~.1~~~~~3~~~~~2~~~~1.7068,~~~1.2678,~~~~1.1613,~~~1.1148,~~~~1.0889,~~~1.0612,~~~1.0377,
\\
~~~~~~~~~~~~~~~~~~~~~~~~~~~~~~~~~~.77462~~~~.80678~~~~~.82742~~~~.84873~~~~~.86583~~~~.89102~~~.92113
\\
\\
~1~~~~~0~~~~~~~9~~~~~1~~~~~1~~~~2.035~~~~~1.6116~~~~~1.5163~~~~1.5028~~~~~1.5511~~~~~~%
$-$~~~~~~~~~$-$ \\
\\
~1~~~~~0~~~~~~~4~~~~~1~~~~~1~~~14.8239~~~~~~~$-$~~~~~~~~~~$-$~~~~~~~~~~$-$%
~~~~~~~~~~~$-$~~~~~~~~~$-$~~~~~~~~~$-$ \\
\\
~1~~~~.1~~~~~~~1~~~~~1~~~~~1~~~~~~~$-$~~~~~~~~~~~$-$~~~~~~~~~~$-$~~~~~~~~~~$%
- $~~~~~~~~~~~$-$~~~~~~~~~$-$~~~~~~~~~$-$ \\
\\
~1~~~~~1~~~~~~~1~~~~~1~~~~~1~~~~1.3104,~~~1.1547,~~~~1.0934,~~~1.0627,~~~~1.045,~~~~~1.0265,~~1.0124,
\\
~~~~~~~~~~~~~~~~~~~~~~~~~~~~~~~~~~~1~~~~~~~~~~~1~~~~~~~~~~~~1~~~~~~~~~~1~~~~~~~~~~~~1~~~~~~~~~~~1~~~~~~~~~1
\\
\\
~4~~~~~0~~~~~~~5~~~~.1~~~~~4~~~~~~~$-$~~~~~~~~~~~$-$~~~~~~~~~~~$-$~~~~~~~~~$%
- $~~~~~~~~~~~$-$~~~~~~~~~$-$~~~~~~~~~$-$ \\
\\
~1~~~~~0~~~~~~~1~~~~~1~~~~~9~~~~2.25,~~~~~~2.5336,~~~~2.2120,~~~1.9675,~~~~1.7967,~~~1.5829,~~1.3756,
\\
~~~~~~~~~~~~~~~~~~~~~~~~~~~~~~~~~~1.5963~~~~1.1725~~~~~1.1045~~~~1.0752~~~~~1.0588~~~~1.041~~~~1.0256
\\
\\
~1~~~~~0~~~~~~~1~~~~~5~~~~~1~~~~~~~$-$~~~~~~~~~~~$-$~~~~~~~~~~$-$~~~~~~~~~~$%
- $~~~~~~~~~~~$-$~~~~~~~~~$-$~~~~~~~~~$-$ \\
\\
~1~~~~~0~~~~~~~1~~~~.1~~~~~1~~~~.90297,~~~.87007,~~~~.86041,~~~.85671,~~~~.85235,~~~~~~%
$-$~~~~~~~~~$-$ \\
~~~~~~~~~~~~~~~~~~~~~~~~~~~~~~~~~~.10733~~~~.36226~~~~~.54185~~~~.66031~~~~.74455
\\
\\ \hline\hline
\end{tabular}
\end{center}

From the definition of the functions in equation (30) it is clear that the
parameters $a,\xi _{0},\zeta _{0}$ are always positive while $\eta
_{0},\Theta _{0}$ can have positive as well as negative values. From the
Table we see that as we decrease the value of $a$, keeping the other
parameters fixed, then the formation of a naked singularity is more probable
in higher dimensions. The situation for $\zeta _{0}$ is similar. But for $%
\xi _{0}$ and $\eta _{0}$ the condition is reversed: black hole formation
becomes more probable as we increase the dimension of the space-time if we
decrease the value of $\xi _{0}$ (or $\eta _{0}$) and keep the other
constants fixed. However, no such conclusions can be drawn about the
variation of $\Theta _{0}$.

\section{{\protect\normalsize \textbf{Strength of the naked singularity}}}

A singularity is called \textit{gravitationally strong,} or simply \textit{%
strong,} if it destroys by crushing or tidally stretching to zero
volume all objects that fall into it; it is called \textit{weak}
if no object that falls into the singularity is destroyed in this
way. A precise characterization of Tipler strong [24]
singularities has been given by Clarke and Krolak [25], who
proposed the strong focusing condition. A sufficient condition
for a strong curvature singularity is that, for at least one non
space-like geodesic with affine parameter $\lambda \rightarrow 0$
on approach to the singularity, we must have
\begin{equation}
\begin{array}{c}
lim \\
\lambda \rightarrow 0 \\
\end{array}%
\begin{array}{c}
\lambda ^{2}R_{ij}K^{i}K^{j}>0 \\
\\
\end{array}%
\end{equation}%
where $K^{i}=\frac{dx^{i}}{d\lambda }$ is the tangent vector to the radial
null geodesic. \newline

Our purpose here to investigate the above condition along future-directed
radial null geodesics that emanate from the naked singularity. Now equation
(35) can be expressed as (using L'H\^{o}pital's rule)
\begin{equation}
\begin{array}{c}
lim \\
\lambda \rightarrow 0 \\
\end{array}%
\begin{array}{c}
\lambda ^{2}R_{ij}K^{i}K^{j} \\
\\
\end{array}%
\begin{array}{c}
=\frac{n\zeta _{0}(H_{0}-\eta _{0}X_{0})(\xi _{0}-(n+1)\eta _{0})}{%
2X_{0}^{^{n}}\left( N_{0}+\eta _{0}\sqrt{\frac{\zeta _{0}}{X_{0}^{n-1}}}%
\right) ^{2}} \\
\end{array}%
\end{equation}%
where $H_{0}=H(X_{0},0),N_{0}=N(X_{0},0)$.\newline

The singularity is gravitationally strong in the sense of Tipler if

\[
\xi _{0}-(n+1)\eta _{0}>max\left\{ 0,-\frac{(n+1)\Theta _{0}}{X_{0}^{\frac{%
n+1}{2}}}\right\}
\]%
or
\[
\xi _{0}-(n+1)\eta _{0}<min\left\{ 0,-\frac{(n+1)\Theta _{0}}{X_{0}^{\frac{%
n+1}{2}}}\right\}
\]

If the above condition is not satisfied for the values of the parameters
then $%
\begin{array}{clll}
lim &  &  &  \\
\lambda \rightarrow 0 &  &  &
\end{array}%
\begin{array}{c}
\lambda ^{2}R_{ij}K^{i}K^{j}\leq 0{}%
\end{array}%
$ and the singularity may or may not be Tipler strong.

\section{{\protect\normalsize \textbf{Discussion and Concluding Remarks}}}

In this paper we have studied gravitational collapse in $(n+2)$-dimensional
space-time using a higher-dimensional generalization of the quasi-spherical
Szekeres metrics with non-zero cosmological constant. We have examined the
local nature of the central shell-focusing singularity by a comparative
study of the time of the formation of trapped surface and the time of
formation of the central shell-focusing singularity. If we assume the
initial density gradient falls off rapidly and vanishes at $r=0,$ then naked
singularity formation is possible only up to space-time dimension five.
However, if we drop the above restriction on the initial density
distribution then the Cosmic Censorship Conjecture may be violated in any
dimension ($n\geq 2$). Thus we deduce that the nature of the central
singularity depends sensitively in these metrics on the choice of the
initial data (particularly on the choice of initial density profile). This
is also confirmed by our geodesic study following the approach of Joshi and
Dwivedi [23] where we have shown numerically that the nature of singularity
depends on the value of the defining parameters at $r=0$. Finally, we have
examined the strength of the naked singularity using the criterion
introduced by Tipler [24]. We found that the naked singularity will be a
strong curvature singularity depending on the appropriate choice of the
value of the parameters at $r=0$. These investigations provide some insights
into the phenomenon of gravitational collapse in a situation without any
imposed Killing symmetries. However, the collapses are special in other
senses which permit exact solutions to be found. In particular, there is an
absence of gravitational radiation in these space-times [27]. An
investigation of its role is a challenge for future analytic and
computational investigations.\newline

\textbf{Acknowledgement:}\newline

One of the authors (U.D) thanks CSIR (Govt. of India) for the award of a
Senior Research Fellowship.\newline

\textbf{References:}\newline
\newline
$[1]$ P. S. Joshi and I. H. Dwivedi, \textit{Commun. Math. Phys.} \textbf{166%
} 117 (1994).\newline
$[2]$ P. S. Joshi and I. H. Dwivedi, \textit{Class. Quantum Grav.} \textbf{16%
} 41 (1999).\newline
$[3]$ K. Lake, \textit{Phys. Rev. Lett.} \textbf{68} 3129 (1992).\newline
$[4]$ A. Ori and T. Piran, \textit{Phys. Rev. Lett.} \textbf{59} 2137 (1987).%
\newline
$[5]$ T. Harada, \textit{Phys. Rev. D} \textbf{58} 104015 (1998).\newline
$[6]$ P.S. Joshi, \textit{Global Aspects in Gravitation and Cosmology, }%
(Oxford Univ. Press, Oxford, 1993).\newline
$[7]$ A. Banerjee, U.Debnath and S. Chakraborty, \textit{gr-qc}/0211099
(2002)(accepted in \textit{Int. J. Mod. Phys. D}) .\newline
$[8]$ U. Debnath and S. Chakraborty, \textit{gr-qc}/0211102 (2002) .\newline
$[9]$ R. Goswami and P.S. Joshi, \textit{gr-qc}/02112097 (2002) .\newline
$[10]$ S. Schoen and S. T. Yau, \textit{Commun. Math. Phys.} \textbf{90} 575
(1983).\newline
$[11]$ K. S. Thorne, in \textit{Magic Without Magic}: \textit{John Archibald
Wheeler}, Ed. Klauder J (San Francisco: W. H. Freeman and Co. 1972).\newline
$[12]$ S. L. Shapiro and S. A. Teukolsky, \textit{Phys. Rev. Lett.} \textbf{%
66} 994 (1991).\newline
$[13]$ T. Nakamura, M. Shibata and K.I. Nakao, \textit{Prog. Theor. Phys.}
\textbf{89} 821 (1993) .\newline
$[14]$ C. Barrabes, W. Israel and P. S. Letelier, \textit{Phys. Lett. A}
\textbf{160} 41 (1991); M. A. Pelath, K. P. Tod and R. M. Wald, \textit{%
Class. Quantum Grav.} \textbf{15} 3917 (1998).\newline
$[15]$ T. Harada, H. Iguchi and K.I. Nakao, \textit{Phys. Rev. D} \textbf{58}
041502 (1998) .\newline
$[16]$ H. Iguchi, T. Harada and K.I. Nakao, \textit{Prog. Theor. Phys.}
\textbf{101} 1235 (1999); \textit{Prog. Theor. Phys.} \textbf{103} 53 (2000).%
\newline
$[17]$ P. Szekeres, \textit{Commun. Math. Phys.} \textbf{41} 55 (1975).%
\newline
$[18]$ P. Szekeres, \textit{Phys. Rev. D} \textbf{12} 2941 (1975).\newline
$[19]$ P. S. Joshi and A. Krolak, \textit{Class. Quantum Grav.} \textbf{13}
3069 (1996).\newline
$[20]$ S. S. Deshingkar, S. Jhingan and P. S. Joshi, \textit{Gen. Rel. Grav.}
\textbf{30} 1477 (1998).\newline
$[21]$ S. M. C. V. Goncalves, \textit{Class. Quantum Grav.} \textbf{18} 4517
(2001).\newline
$[22]$ S. Chakraborty and U. Debnath , \textit{gr-qc}/0304072 (2003).\newline
$[23]$ P. S. Joshi and I. H. Dwivedi, \textit{Phys. Rev. D} \textbf{47} 5357
(1993).\newline
$[24]$ F. J. Tipler, \textit{Phys. Lett. A} \textbf{64} 8 (1987).\newline
$[25]$ C. J. S. Clarke and A. Krolak, \textit{J. Geom. Phys.} \textbf{2} 127
(1986).\newline
$[26]$ J.D. Barrow and J. Stein-Schabes, \textit{Phys. Lett. A} \textbf{103}%
, 315 (1984).\newline
$[27]$ W. B. Bonnor, \textit{Comm. Math.
Phys}. \textbf{51}, 191 (1976)

\end{document}